\documentclass[review]{elsarticle}
\usepackage{bm,amssymb,amsmath,mathrsfs}
\usepackage{amsthm}
\usepackage{lineno}
\usepackage[usenames]{color}

\usepackage{dcolumn}% Align table columns on decimal point

\theoremstyle{plain}

\newtheorem*{theorem*}{Theorem}
\begin{document}

\journal{(internal report CC25-1)}
%\journal{}

\begin{frontmatter}

\title{Polarized electron bunch refresh rates in an electron storage ring}

\author[cc]{S.~R.~Mane}
\ead{srmane001@gmail.com}
\address[cc]{Convergent Computing Inc., P.~O.~Box 561, Shoreham, NY 11786, USA}

\begin{abstract}
  When polarized electron bunches are injected and circulated in a high-energy storage ring,
  the polarization of the bunches relaxes to the asymptotic value of the radiative polarization, caused by the synchrotron radiation.
  Hence the bunches must be refreshed periodically, to maintain a predetermined time-averaged value of the bunch polarization.
  In general, the refresh rates of the ``up'' and ``down'' polarization bunches are different.
  We suggest an alternative policy.
  We point out that the total bunch refresh rate is almost independent of the asymptotic level of the radiative polarization.
  We also note that the true goal of so-called ``spin matching'' is to maximize the buildup time constant (not the asymptotic level) of the radiative polarization.
  We suggest a scheme to equalize the refresh rates of the ``up'' and ``down'' polarization bunches,
  which may be (i) helpful for accelerator operations, and also (ii) reduce systematic errors in HEP experiments.
\end{abstract}

\end{frontmatter}

\newpage
\section{Introduction}\label{sec:intro}
The electron-ion collider (EIC) project has been approved, to be sited at Brookhaven National Lab (BNL).
Spin polarized bunches of electrons will circulate in an electron storage ring (ESR)
and collide at selected locations (interaction points) with bunches of ions stored in a separate counterrotating ring.
The technical details of the project can be found in the Conceptual Design Report (CDR) \cite{CDR} of the EIC project.
Electron bunches (from a polarized electron source) with ``up'' and ``down'' vertical polarizations will be injected and stored into the ESR.
During storage, the polarization levels of the bunches relax to an asymptotic value due to the radiative polarization generated by the synchrotron radiation \cite{ST,DK73}.
Hence the bunches must be refreshed periodically, to maintain a predetermined time-averaged value of the bunch polarization.

For example (taking numbers from \cite{CDR}),
if the polarization level at injection is $85\%$ and the desired time-averaged polarization is $70\%$,
then a bunch must be refreshed when its polarization level reaches $55\%$ (approximately).
In the proposed design of the ESR, the asymptotic radiative polarization points vertically up.
For the sake of argument, suppose that the asymptotic level of the radiative polarization is $35\%$.
The polarization of the ``up'' bunches relaxes from $+85\%$ to $+35\%$ (a difference of $50\%$)
but the polarization of the ``down'' bunches relaxes from $-85\%$ to $+35\%$ (a difference of $120\%$).
For this reason, the ``down'' bunches must be refreshed more frequently than the ``up'' bunches.

We suggest a scheme to equalize the refresh rates of the ``up'' and ``down'' polarization bunches.
This may be (i) helpful for accelerator operations, and also (ii) reduce systematic errors in HEP experiments.
To be clear, the analysis below is applicable for high-energy electron storage rings in general, and is not specific to the EIC project at BNL.
It will be shown that the total refresh rate of all the bunches is insensitive (almost independent) of the asymptotic value of the radiative polarization.
We also note that the true goal of so-called ``spin matching'' is to maximize the buildup time constant (not the asymptotic level) of the radiative polarization.

\section{Bunch refresh rates}\label{sec:refresh}
\subsection{Parameter definitions and full formulas}
First let us introduce the following parameters.
Here and below, `DK' referes to `Derbenev and Kondratenko'
(see \cite{DK73} for explicit formulas for the asymptotic level and buildup time of the radiative polarization in high-energy storage rings).
Define the following:
\begin{enumerate}
\item
  $P_0=$ polarization level at injection.
\item
  $P_{\min}=$ minimum acceptable polarization level (then refresh the bunch).
\item
  $P_{DK}=$ asymptotic level of radiative polarization.
\item
  $\tau_{DK}=$ time constant for decay of the radiative polarization to its asymptotic level.
\item
  $t_1=$ time to refresh bunch with ``up'' polarization
\item
  $t_2=$ time to refresh bunch with ``down'' polarization
\item
  $R=1/t_1+1/t_2=$ total refresh rate of ``up'' and ``down'' bunches
\end{enumerate}
For the bunches with ``up'' polarization, the polarization level $P_u(t)$ at time $t$ is ($P_u(0) = P_0$)
\begin{equation}  
\label{eq:Pt_u}   
P_u(t) = P_{DK} + (P_0 - P_{DK})e^{-t/\tau_{DK}} \,.
\end{equation}
For the bunches with ``down'' polarization, the polarization level $P_d(t)$ at time $t$ is ($P_d(0) = -P_0$)
\begin{equation}  
\label{eq:Pt_d}   
P_d(t) = P_{DK} - (P_0 + P_{DK})e^{-t/\tau_{DK}} \,.
\end{equation}
By definition $P_u(t_1) = P_{\min}$ at $t=t_1$, hence
\begin{equation}  
\label{eq:Pt1}
\begin{split}
  P_{\min} &= P_{DK} + (P_0 - P_{DK})e^{-t_1/\tau_{DK}}
  \\
  P_{\min} - P_{DK} &= (P_0 - P_{DK})e^{-t_1/\tau_{DK}}
  \\
  e^{t_1/\tau_{DK}} &= \frac{P_0 - P_{DK}}{P_{\min} - P_{DK}}
  \\
  t_1 &= \tau_{DK} \ln\Bigl(\frac{P_0 - P_{DK}}{P_{\min} - P_{DK}}\Bigr) \,.
\end{split}
\end{equation}
Also by definition $P_d(t_1) = -P_{\min}$ at $t=t_2$, hence
\begin{equation}  
\label{eq:Pt2}
\begin{split}
  -P_{\min} &= P_{DK} - (P_0 + P_{DK})e^{-t_2/\tau_{DK}}
  \\
  P_{\min} + P_{DK} &= (P_0 + P_{DK})e^{-t_2/\tau_{DK}}
  \\
  e^{t_2/\tau_{DK}} &= \frac{P_0 + P_{DK}}{P_{\min} + P_{DK}}
  \\
  t_2 &= \tau_{DK} \ln\Bigl(\frac{P_0 + P_{DK}}{P_{\min} + P_{DK}}\Bigr) \,.
\end{split}
\end{equation}
The refresh rate for the ``up'' polarization bunches is $R_1=1/t_1$, given by
\begin{equation}
\label{eq:r1}  
R_1 = \frac{1}{t_1} = \frac{1}{\tau_{DK}} \frac{1}{\displaystyle \ln\Bigl(\frac{P_0 - P_{DK}}{P_{\min} - P_{DK}}\Bigr)} \,.
\end{equation}
The refresh rate for the ``down'' polarization bunches is $R_2=1/t_2$, given by
\begin{equation}
\label{eq:r2}  
R_2 = \frac{1}{t_2} = \frac{1}{\tau_{DK}} \frac{1}{\displaystyle \ln\Bigl(\frac{P_0 + P_{DK}}{P_{\min} + P_{DK}}\Bigr)} \,.
\end{equation}
The total bunch refresh rate is $R=R_1+R_2=1/t_1+1/t_2$, given by
\begin{equation}
\label{eq:R}  
R = R_1+R_2 = \frac{1}{\tau_{DK}}
\biggl[ \frac{1}{\displaystyle \ln\Bigl(\frac{P_0 - P_{DK}}{P_{\min} - P_{DK}}\Bigr)} +\frac{1}{\displaystyle \ln\Bigl(\frac{P_0 + P_{DK}}{P_{\min} + P_{DK}}\Bigr)} \biggr] \,.
\end{equation}

\subsection{Comments}\label{sec:comm}
The following symmetry is a self-consistency check.
The expression for $t_2$ is the same as that for $t_1$, with $P_{DK}$ replaced by $-P_{DK}$.
This can be easily understood as follows.
The decay of the polarization from an initial value $-P_0$ to an asymptotic value $P_{DK}$
is the same as the decay from an initial value $P_0$ to an asymptotic value $-P_{DK}$:
just flip one's perspective of the ring upside-down.

Next, let us distinguish between the bunch \emph{refresh} rate (see eq.~\eqref{eq:R}) and
the bunch \emph{replacement} rate.
If there are totally $N_b$ bunches in the ring, half with ``up'' and half with ``down'' polarization, then the total number of electrons which must be replaced per unit time is
$(N_b/2)(R_1+R_2)$.
This is the bunch replacement rate.
We multiply by the electron charge to obtain the beam current which must be delivered by the polarized electron source.
Since $N_b$ is a constant, the bunch replacement rate is not relevant to the analysis in this note.

\subsection{Linearized approximation}\label{sec:lin}
To simplify the formulas, assume the buildup of the radiative polarization is slow, i.e.~$\tau_{DK}\gg t_1$ and $\tau_{DK}\gg t_2$.
Hence let us linearize $e^{-t/\tau_{DK}} \simeq 1 - t/\tau_{DK}$.
(This calculation was suggested by K.~Yokoya \cite{Yokprivcomm}.)
Approximate for $P_u(t)$ as follows
\begin{equation}  
\label{eq:Pt_u_lin}
\begin{split}
P_u(t) &= P_{DK} + (P_0 - P_{DK})e^{-t/\tau_{DK}} 
\\
&\simeq P_{DK} + (P_0 - P_{DK})\Bigl(1 -\frac{t}{\tau_{DK}}\Bigr)
\\
&= P_0 - (P_0 - P_{DK})\frac{t}{\tau_{DK}} \,.
\end{split}
\end{equation}
Next approximate for $P_d(t)$ as follows
\begin{equation}  
\label{eq:Pt_d_lin}
\begin{split}
P_d(t) &= P_{DK} - (P_0 + P_{DK})e^{-t/\tau_{DK}} 
\\
&\simeq P_{DK} - (P_0 + P_{DK})\Bigl(1 -\frac{t}{\tau_{DK}}\Bigr)
\\
&= -P_0 + (P_0 + P_{DK})\frac{t}{\tau_{DK}} \,.
\end{split}
\end{equation}
By definition $P_u(t_1) = P_{\min}$ at $t=t_1$, hence
\begin{equation}  
\label{eq:Pt1_lin}
\begin{split}
  P_{\min} &\simeq P_0 - (P_0 - P_{DK})\frac{t_1}{\tau_{DK}}
  \\
  (P_0 - P_{DK})\frac{t_1}{\tau_{DK}} &\simeq P_0 - P_{\min}
  \\
  t_1 &\simeq \tau_{DK} \frac{P_0 - P_{\min}}{P_0 - P_{DK}} \,.
\end{split}
\end{equation}
Also by definition $P_d(t_2) = -P_{\min}$ at $t=t_2$, hence
\begin{equation}  
\label{eq:Pt2_lin}
\begin{split}
  -P_{\min} &\simeq -P_0 + (P_0 + P_{DK})\frac{t_2}{\tau_{DK}}
  \\
  (P_0 + P_{DK})\frac{t_2}{\tau_{DK}} &\simeq P_0 - P_{\min}
  \\
  t_2 &\simeq \tau_{DK} \frac{P_0 - P_{\min}}{P_0 + P_{DK}} \,.
\end{split}
\end{equation}
Then the refresh rate of the ``up'' polarization bunches is
\begin{equation}
\label{eq:r1_lin}  
R_1 = \frac{1}{t_1} \simeq \frac{1}{\tau_{DK}} \frac{P_0 - P_{DK}}{P_0 - P_{\min}} \,.
\end{equation}
Also the refresh rate of the ``down'' polarization bunches is
\begin{equation}
\label{eq:r2_lin}  
R_2 = \frac{1}{t_2} \simeq \frac{1}{\tau_{DK}} \frac{P_0 + P_{DK}}{P_0 - P_{\min}} \,.
\end{equation}
The total bunch refresh rate is
\begin{equation}
\begin{split}
R = R_1+R_2 &\simeq \frac{1}{\tau_{DK}} \biggl[ \frac{P_0 - P_{DK}}{P_0 - P_{\min}} + \frac{P_0 + P_{DK}}{P_0 - P_{\min}} \biggr]
\\
&= \frac{1}{\tau_{DK}} \frac{2P_0}{P_0 - P_{\min}} \,.
\end{split}
\end{equation}
\emph{The total bunch refresh rate is independent of the value of the asymptotic polarization level $P_{DK}$.}

\subsection{Numbers}
Let us try a few numbers.
Note that $P_{DK}$ and $\tau_{DK}$ can be varied independently.
Hence we hold the values of $\tau_{DK}$, $P_0$ and $P_{\min}$ fixed and vary only $P_{DK}$.
Then we only need to compute the value of
\begin{equation}
X = \frac{1}{\displaystyle \ln\Bigl(\frac{P_0 - P_{DK}}{P_{\min} - P_{DK}}\Bigr)} +\frac{1}{\displaystyle \ln\Bigl(\frac{P_0 + P_{DK}}{P_{\min} + P_{DK}}\Bigr)} \,.
\end{equation}
From (\cite{CDR}, p.~437 ff.),
suppose the injected polarization is $85\%$ and the desired time-averaged polarization is $70\%$.
Then set $P_{\min} = 55\%$ (approximately).
The values of $X$, $t_1/\tau_{DK}$ and $t_2/\tau_{DK}$, using $P_0=85\%$ and $P_{\min}=55\%$,
are tabulated in Table \ref{tb:numbers}.
Note the following.
\begin{enumerate}
\item
  As is necessary,
  \\(i) $t_1/\tau_{DK}$ increases as $P_{DK}$ increases (longer time interval to refresh the bunch),
  \\(ii) $t_2/\tau_{DK}$ increases as $P_{DK}$ increases (shorter time interval to refresh the bunch).
\item
  Observe that $t_1/\tau_{DK} > 1$ for $P_{DK} > 40\%$, which indicates that a linearized approximation is not good.
\item
  Nevertheless, observe that the value of $X$ is almost constant, as a function of $P_{DK}$.
\item
  The total bunch refresh rate is insensitive to $P_{DK}$, the asymptotic value of the radiative polarization.
\item
  The total bunch refresh rate $R$ depends almost only on $\tau_{DK}$, and weakly on $P_{DK}$.
\end{enumerate}

\newpage
\section{Alternative analysis}\label{sec:georg}
Georg Hoffstaetter \cite{GHprivcomm} kindly informed the author of the following details.
In the BNL ESR, it is desired to have the same \emph{time-averaged polarization level} for the
``up'' and ``down'' polarization bunches, not the same value of $P_{\min}$.
Hence let us rework the analysis.
Let the time-averaged polarization level be $P_a$.
To avoid confusion, denote the refresh time for the ``up'' and ``down'' polarization bunches, by $t_3$ and $t_4$, respectively.
Then for the ``up'' polarization bunches (see eq.~\eqref{eq:Pt_u})
\begin{equation}
\begin{split}
P_a = \frac{1}{t_3}\int_0^{t_3} P_u(t)\,dt
&= \frac{1}{t_3}\int_0^{t_3} \biggl[P_{DK} + (P_0 - P_{DK})e^{-t/\tau_{DK}}\biggr]\,dt
\\
&= P_{DK} + (P_0 - P_{DK})(1-e^{-t_3/\tau_{DK}})\frac{\tau_{DK}}{t_3} \,.
\end{split}
\end{equation}
Hence
\begin{equation}
\frac{1-e^{-t_3/\tau_{DK}}}{t_3/\tau_{DK}} = \frac{P_a - P_{DK}}{P_0 - P_{DK}} \,.
\end{equation}
Next for the ``down'' polarization bunches (see eq.~\eqref{eq:Pt_d})
\begin{equation}
\begin{split}
P_a = -\frac{1}{t_4}\int_0^{t_4} P_d(t)\,dt
&= -\frac{1}{t_4}\int_0^{t_4} \biggl[P_{DK} - (P_0 + P_{DK})e^{-t/\tau_{DK}}\biggr]\,dt
\\
&= -P_{DK} + (P_0 + P_{DK})(1-e^{-t_4/\tau_{DK}})\frac{\tau_{DK}}{t_4} \,.
\end{split}
\end{equation}
Hence
\begin{equation}
  \frac{1-e^{-t_4/\tau_{DK}}}{t_4/\tau_{DK}} = \frac{P_a + P_{DK}}{P_0 + P_{DK}} \,.
\end{equation}
Observe that the solution for $t_4$ is obtained by replacing $P_{DK}$ by $-P_{DK}$
in the formula for $t_3$.
The values of $t_3$ and $t_4$ must be calculated numerically.
Let us fix values for $P_0(=85\%)$ and $P_a(=70\%)$ and sweep the value of $P_{DK}$ and compute $t_3/\tau_{DK}$, $t_4/\tau_{DK}$ and $Y = \tau_{DK}(1/t_3+1/t_4)$.
The results are tabulated in Table \ref{tb:numbers2}.
The value of $Y$ is \emph{even more insensitive} to the value of $P_{DK}$.
(Here $P_{\min\,3}$ and $P_{\min\,4}$ are the polarization levels of the
``up'' and ``down'' polarization bunches at time $t_3$ and $t_4$, respectively.
They are displayed just for reference and play no role in the analysis.)

\section{Proposed scheme to equalize bunch refresh rates}\label{sec:scheme}
If the asymptotic value of the radiative polarization is zero,
the polarization in both the ``up'' and ``down'' bunches will decay to zero.
Then the refresh rates of the ``up'' and ``down'' polarization bunches are equal.
We can set the asymptotic value of the radiative polarization to zero as follows.
We insert a pair of diametrically opposed Siberian Snakes with orthogonal spin rotation axes in the ring.
Consult \cite{MSY1} for details about Siberian Snakes in high-energy storage rings
(and consult \cite{MSY2} more generally for a review of spin dynamics in high-energy particle accelerators).
This will cause the polarization direction to be vertically up in one half of the ring and down in the other half.
The asymptotic value of the radiative polarization $P_{DK}$ cancels to zero.

Note that just because $P_{DK}=0$ does not mean that the time constant $\tau_{DK}$ can be ignored.
The ring optics must be adjusted to maximize the value of $\tau_{DK}$, to minimize the bunch refresh rates.
So-called ``spin matching'' is required to accomplish this.
See for example \cite{BuonSteffen} for the spin matching of the HERA minirotator in the lepton ring of HERA.
More modern techniques of spin matching are currently under investigation, for example see \cite{SH_BAGELS2},
which lists specific applications to the current design of the ESR of the EIC project at BNL (but without Siberian Snakes in the ESR lattice).

Note also that with diametrically opposed Siberian Snakes with orthogonal spin rotation axes, the spin tune equals $\frac12$ at all beam energies, which is also helpful.
This may also aid with machine operations.
Detailed conclusions will require the analysis of specific accelerator lattices.

In general, Siberian Snakes composed of dipole bends may entail large transverse orbit excursions, which may not be practical to accomodate.
(For example, it is not practical for the ESR of the EIC project at BNL at its lower operating energy of 5 GeV.)
Siberian Snakes composed of solenoids have no transverse orbit excursions, but the integrated magnetic field is proportional to the beam momentum,
and may be unacceptably large at very high beam energies.
Hence there is a tradeoff: solenoid Siberian Snakes may be practical at lower beam energies and dipole Siberian Snakes may be practical at higher beam energies.
The design choice depends on the operating energy range of the electron storage ring.
In the case of the ESR of the EIC project at BNL, where the operating energy range is $5-18$ GeV, solenoid Siberian Snakes are practical.
Solenoid Siberian Snakes with longitudinal spin rotation axes were operated at the
Amsterdam Pulse Stretcher (AmPS) \cite{AmPS} and the MIT-Bates South Hall Ring (SHR) \cite{SHR}
and their design, including matching to the lattice optics, is known.
The design of a radial axis solenoid Siberian Snake is listed in (\cite{CDR}, Fig.~5.13), reproduced schematically below
$$  
  \textrm{Arc} \;-\; \textrm{Sol1} \;-\; \textrm{Bend1} \;-\; \textrm{Sol2} \;-\; \textrm{Bend2} \;-\;
  \textrm{Bend2} \;-\; (-\textrm{Sol2}) \;-\; \textrm{Bend1} \;-\; (-\textrm{Sol1}) \;-\; \textrm{Arc}
$$
Here ``Sol1'' and ``Sol2'' denote ``solenoid modules'' (solenoids, quadrupoles and skew quadrupoles to match the lattice optics),
and  ``$-$Sol1'' and ``$-$Sol2'' denote the same modules, with the solenoid fields reversed.
The bending sections may also include quadrupoles.
The spin rotations are symmetric around the midpoint for the bending sections and antisymmetric for the solenoids.

\section{Conclusion}\label{sec:conc}
It has been shown that for electron storage rings where polarized electron bunches are injected into the ring,
the total bunch refresh rate, to maintain a predetermined time-averaged value of the bunch polarization,
depends only weakly on the asymptotic level of the radiative polarization.
The true goal of so-called ``spin matching'' is to maximize the buildup time constant (not the asymptotic level) of the radiative polarization.
This note suggests a scheme to make the asymptotic level of the radiative polarization equal zero.
If the asymptotic value of the radiative polarization is zero, the polarizations of both the ``up'' and ``down'' bunches will decay to zero.
Then the refresh rates of the ``up'' and ``down'' bunches will be equal.
Also, the ``sawtooth'' of the time-dependence of the polarizations will be identical for the ``up'' and ``down'' bunches,
which is helpful to reduce systematic errors in HEP data analysis of the electron-ion collisions.
The proposed scheme employs diametrically opposed Siberian Snakes with orthogonal spin rotation axes.
Then the spin tune equals $\frac12$ at all beam energies, which is also helpful.

\section*{Acknowledgements}
The author thanks K.~Yokoya for helpful discussions,
especially to show that in the linearized approximation,
the total bunch refresh rate is independent of the asymptotic value of the radiative polarization.
The author also thanks Georg Hoffstaetter for kindly supplying pertinent details about the proposed operation of the BNL ESR.

%\newpage
\bibliographystyle{amsplain}

\newpage
\begin{table}[!htb]
\centering
\smallskip
\begin{tabular}[width=0.75\textwidth]{rlll}
\hline
    $P_{DK}$ (\%) & $X$ & $t_1/\tau_{DK}$ & $t_2/\tau_{DK}$ \\
\hline
\hline
0	&	4.5943	&	0.4353	&	0.4353	\\
5	&	4.5939	&	0.4700	&	0.4055	\\
10	&	4.5927	&	0.5108	&	0.3795	\\
15	&	4.5906	&	0.5596	&	0.3567	\\
20	&	4.5874	&	0.6190	&	0.3364	\\
25	&	4.5829	&	0.6931	&	0.3185	\\
30	&	4.5765	&	0.7885	&	0.3023	\\
35	&	4.5674	&	0.9163	&	0.2877	\\
40	&	4.5541	&	1.0986	&	0.2744	\\
45	&	4.5328	&	1.3863	&	0.2624	\\
50	&	4.4929	&	1.9459	&	0.2513	\\
\hline
\end{tabular}
\caption{\small \label{tb:numbers} Values of selected parameters as a function of the asymptotic radiative polarization level $P_{DK}$.}
\end{table}

\newpage
\begin{table}[!htb]
\centering
\smallskip
\begin{tabular}[width=0.75\textwidth]{rlllllllll}
\hline
    $P_{DK}$ (\%) & $Y$ & $t_3/\tau_{DK}$ & $t_4/\tau_{DK}$ & $P_{\min\,3}$ & $P_{\min\,4}$ \\
\hline
\hline
0	&	4.9783	&	0.4017	&	0.4017	&	56.8779	&	56.8779	\\
5	&	4.9782	&	0.4307	&	0.3764	&	57.0035	&	56.7672	\\
10	&	4.9779	&	0.4642	&	0.3541	&	57.1472	&	56.6687	\\
15	&	4.9774	&	0.5034	&	0.3343	&	57.3131	&	56.5807	\\
20	&	4.9767	&	0.5499	&	0.3167	&	57.5070	&	56.5015	\\
25	&	4.9757	&	0.6059	&	0.3007	&	57.7363	&	56.4298	\\
30	&	4.9743	&	0.6747	&	0.2864	&	58.0120	&	56.3647	\\
35	&	4.9725	&	0.7614	&	0.2733	&	58.3498	&	56.3053	\\
40	&	4.9701	&	0.8742	&	0.2614	&	58.7735	&	56.2508	\\
45	&	4.9667	&	1.0272	&	0.2504	&	59.3207	&	56.2007	\\
50	&	4.9619	&	1.2472	&	0.2404	&	60.0556	&	56.1545	\\
\hline
\end{tabular}
\caption{\small \label{tb:numbers2} Values of selected parameters as a function of the asymptotic radiative polarization level $P_{DK}$.}
\end{table}

\end{document}